%% file: main.tex
\documentclass[sigconf]{acmart}
\AtBeginDocument{%
  }

\setcopyright{acmlicensed}
\copyrightyear{2025}
\acmYear{2025}
\acmDOI{XXXXXXX.XXXXXXX}
\acmConference[EASE 2025]{The 29th International Conference on Evaluation and Assessment in Software Engineering}{17–20 June, 2025}{Istanbul, Türkiye}
\acmISBN{978-1-4503-XXXX-X/18/06}

\usepackage{hyperref}
\usepackage{array}
\usepackage{makecell, multirow, booktabs}
\usepackage{graphicx}
\usepackage{caption}
\usepackage{subcaption}
\usepackage{packages}
\usepackage{hyperxmp}

\begin{document}

\title[How Do Communities of ML-Enabled Systems Smell? A Cross-Sectional Study on the Prevalence of Community Smells]{How Do Communities of ML-Enabled Systems Smell?\\A Cross-Sectional Study on the Prevalence of Community Smells}

\author{Giusy Annunziata}
\email{gannunziata@unisa.it}
\orcid{0009-0002-0742-7261}
\affiliation{%
  \institution{University of Salerno}
  \city{Salerno}
  \country{Italy}
}

\author{Stefano Lambiase}
\email{slambiase@unisa.it}
\orcid{0000-0002-9933-6203}
\affiliation{%
  \institution{University of Salerno}
  \city{Salerno}
  \country{Italy}
}

\author{Fabio Palomba}
\email{fpalomba@unisa.it}
\orcid{0000-0001-9337-5116}
\affiliation{%
  \institution{University of Salerno}
  \city{Salerno}
  \country{Italy}
}

\author{Gemma Catolino}
\email{gcatolino@unisa.it}
\orcid{0000-0002-4689-3401}
\affiliation{%
  \institution{University of Salerno}
  \city{Salerno}
  \country{Italy}
}

\author{Filomena Ferrucci}
\email{fferrucci@unisa.it}
\orcid{0000-0002-0975-8972}
\affiliation{%
  \institution{University of Salerno}
  \city{Salerno}
  \country{Italy}
}

\renewcommand{\shortauthors}{Annunziata et al.}

\begin{abstract}
Effective software development relies on managing both collaboration and technology, but socio-technical challenges can harm team dynamics and increase technical debt. Although teams working on ML-enabled systems are interdisciplinary, research has largely focused on technical issues, leaving their socio-technical dynamics underexplored.
This study aims to address this gap by examining the prevalence, evolution, and interrelations of ``community smells'', in open-source ML projects. We conducted an empirical study on 188 repositories from the NICHE dataset using the CADOCS tool to identify and analyze community smells. Our analysis focused on their prevalence, interrelations, and temporal variations. We found that certain smells—such as Prima Donna Effects and Sharing Villainy—are more prevalent and fluctuate over time compared to others like Radio Silence or Organizational Skirmish. These insights might provide valuable support for ML project managers in addressing socio-technical issues and improving team coordination.
\end{abstract}


\maketitle

\input{Sections/1.intro}
\input{Sections/2.SoA}
\input{Sections/3.rm}
\input{Sections/4.results}
\input{Sections/5.discussion-implication}
\input{Sections/6.ttv}
\input{Sections/7.conclusion}


\section*{Acknowledgment}
This work has been partially supported by the \textsl{EMELIOT} national research project, which has been funded by the MUR under the PRIN 2020 program (Contract 2020W3A5FY), and \textsl{QUAL-AI} national research projects funded by the EU - NGEU and the MUR under the PRIN 2022 program (Contracts 2022B3BP5S).

\balance
\bibliographystyle{ACM-Reference-Format}
\bibliography{bibliography}

\appendix

\end{document}

%% file: Sections/1.intro.tex
\section{Introduction}
\label{sec:Intro}

The success of a software development project depends not only on the technical aspects of the product but also on effective collaboration and communication within the development team~\cite{tamburri2015social,martini2017revealing}. Social dynamics can directly impact technical processes, influencing both productivity and software quality ~\cite{palombaTechnicalAspectsHow2021}.
However, such social challenges, being abstract factors rooted in communication and collaboration, are inherently complex to detect and quantify~\cite{tamburri2016_architect_role_in_community}.
The literature has investigated ``\textit{Community Smells}'', which are measurable patterns of organizational problems that hinder cooperation and affect project health~\cite{tamburri2015open-source}. These smells often lead to increased social debt, where unresolved social issues accumulate over time, degrading software quality and team cohesion~\cite{palombaTechnicalAspectsHow2021,catolino2021understanding}.

Machine Learning (ML)-enabled systems introduce additional complexity when analyzing social dynamics. By their nature, these systems rely on multidisciplinary teams composed of developers, software engineers, and data scientists, each bringing distinct educational background, perspectives, and expertise~\cite{ozkaya2020really}. While this diversity can foster innovation and improve problem-solving, it can also create social tensions, misalignments, and communication gaps that may exacerbate the emergence of community smells~\cite{Busquim2024interactionSEDS}. 

Despite the critical role of social aspects in the shaping of ML-enabled systems, they remain largely underexplored compared to technical aspects. Most research to date has primarily focused on issues such as technical debt, suboptimal design decisions, or neglected maintenance tasks~\cite{foidl2022data,recupito2024unmasking}, often leaving the social dimension insufficiently explored~\cite{SocioTechnicalAntiPattern}.


An initial investigation in this direction was conducted by \citeauthor{SocioTechnicalAntiPattern}~\cite{SocioTechnicalAntiPattern}, who explored socio-technical anti-patterns identified by the communities of ML-enabled systems, analyzing their causes and potential mitigation strategies.
Building on this, \citeauthor{annunziata2025Uncovering}~\cite{annunziata2025Uncovering} examined the correlation between these anti-patterns and known community smells, providing the first insight into community smells within communities of ML-enabled systems.
Their findings highlight that the interaction between diverse teams and the complex workflows characterizing those projects can exacerbate community smells and social issues, leading to increased social debt and weakening collaboration and productivity. Therefore, analyzing these socio-technical issues is critical to developing strategies that improve team cohesion, mitigate social tensions, and improve the overall health and results of software projects.
\smallskip
\examplebox{Our study aims to deepen the understanding and the examination of the socio-technical well-being of communities developing ML-enabled systems, providing a first investigation of the prevalence and evolution of community smells in real open-source ML-enabled systems.}

\vspace{3mm}
We conducted our study on 188 GitHub ML-enabled open-source projects from the NICHE dataset~\cite{widyasari2023niche}, identifying and detecting 10 distinct community smells, e.g., Prima Donna Effect and Toxic Communication show the adoption of unhealthy and rude communication, superiority, and uncooperative behavior, 
using the CADOCS tool~\cite{voria2022_CADOCS}.
Next, we performed a Cross-Sectional Study to examine the prevalence of each community smell in the analyzed projects and explore correlations between them.
Finally, we analyzed the evolution of community smells over time through a Longitudinal Study, examining each project in 3-month intervals from its inception to 2 years of age.


The results obtained show that Prima Donna Effect (PDE)—which refers to a superiority, disagreement, and uncooperative behavior—is the most prevalent smell, whit a prevalence of 92.6\%. Furthermore, the longitudinal analysis reveals that PDE remains consistently high over time (95-96\%).
In contrast, Radio Silence (RS)—where formal communication barriers between sub-communities cause time loss —is the least frequent, observed in 18,6\% of the cases.
Meanwhile, other community smells tend to fluctuate over time, for example, Unhealthy Interaction (UI)—the adoption of unhealthy behavior among team members—decreased by 24\% to 6\%, suggesting improvements in communication practices. 
Finally, the results suggest a strong correlation between PDE and other community smells, in particular the Organizational Silo Effect (OSE)—which describes isolated communities that communicate with each other only through a few members. 


Our findings offer valuable insights for project managers in ML-enabled projects, equipping them with a deeper understanding of the most common social challenges and practical strategies to address them. Additionally, these results serve as a strong foundation for researchers, providing an initial exploration and quantitative analysis of community smells in the ML-enabled context. By highlighting their evolution during the early stages of software projects, this work sets the stage for future investigations into the interplay between human dynamics and their impact on the technical success of ML-enabled projects.

\smallskip
\noindent \textbf{Structure of the paper.} Section \ref{sec:SoA} presents background and related work on ML-enabled systems and community smells.
Section \ref{sec:rm} outlines the design of the study and the methodology for each research question. 
Section \ref{sec:results} reports the results of the study.
Section \ref{sec:dis_imp} discusses the insights of the paper and practical implications, while Section \ref{sec:ttv} the threats to validity. 
Finally, Section \ref{sec:conclusion} concludes the paper and outlines future investigations.

%% file: Sections/2.SoA.tex
\section{Background and Related Work}
\label{sec:SoA}

In this section, we discuss the background and related work that underpin our study.

\subsection{ML-Enabled Systems}

Machine Learning-enabled systems (ML-Enabled Systems) are a specialized category of software systems that integrate machine learning models to automate and enhance decision-making processes~\cite{Villamizar2022SpecificationMLSystems,martinez22aibasedsoftware}. These systems are increasingly deployed across various domains, including autonomous vehicles and healthcare, leveraging ML algorithms to analyze data, recognize patterns, and continuously improve through learning.
Given the complexity of developing and integrating ML components into a system, both technical and social challenges arise throughout the development of an ML-enabled system. Technical debt is a clear example of a technical challenge, referring to the accumulation of suboptimal design decisions or deferred maintenance tasks that ultimately increase the cost of future system improvements and maintenance ~\cite{zhang2022code}. A significant aspect of technical debt is the presence of code smells and data smells—indicators of underlying issues in software and data management~\cite{li2015systematic,foidl2022data,recupito2024unmasking}. Code smells manifest in various ways and become particularly relevant in the context of ML models, data pipelines, and experimentation workflows. Similarly, data smells—issues related to data collection, storage, and preprocessing—represent a significant source of technical debt, potentially leading to biases, reduced model accuracy, and compromised fairness \cite{foidl2022data}. \citeauthor{recupito2024unmasking}\cite{recupito2024unmasking} further underscores the impact of data smells in AI systems, linking them to challenges such as biased predictions and degraded performance.

Social factors also play a critical role in the accumulation of technical debt in ML-enabled systems. In particular, social challenges are significant due to the \emph{heterogeneous nature} of the teams involved, which often include diverse roles such as data scientists and software engineers. While this diversity can foster innovation, it can also introduce socio-technical issues. Misunderstandings regarding data requirements or deployment constraints can lead to inefficiencies, rework, and increased technical debt, especially when changes to a system component are not effectively communicated across the team~\cite{ML-EnabledSystemsCCDE}.
For instance, \citeauthor{Busquim2024interactionSEDS}~\cite{Busquim2024interactionSEDS} conducted semi-structured interviews with software engineering and data science professionals, revealing challenges such as disparities in technical expertise, ambiguous roles, and inadequate documentation. These challenges exacerbate inefficiencies and technical debt, underscoring the need for improved collaboration and communication strategies~\cite{ML-EnabledSystemsCCDE}.
Similarly, \citeauthor{ML-EnabledSystemsCCDE}~\cite{ML-EnabledSystemsCCDE} investigated collaboration and communication challenges during ML-enabled system development. They found that miscommunication, rooted in the differing technical vocabularies among the various team members, was a primary issue. Such communication gaps hinder the understanding between developers and managers, further complicating project workflows.
In another study, \citeauthor{SocioTechnicalAntiPattern}~\cite{SocioTechnicalAntiPattern} explored socio-technical aspects in ML-enabled systems by analyzing 73 relevant videos from the ML-Ops community.\footnote{ML-Ops Community: \url{https://mlops.community/}} They identified 17 socio-technical anti-patterns, which describe problematic developer behaviors, along with 16 causes and 15 organizational strategies to address them. 
However, the study raises an open question about such anti-patterns, asking whether they are specific to ML-enabled systems or are simply manifestations of socio-technical challenges already investigated in the context of software engineering.

\subsection{Community Smells in the Context of ML-enabled Systems}

Community smells, introduced by \citeauthor{tamburri2015open-source}~\cite{tamburri2015open-source}, are measurable socio-technical patterns that negatively impact team collaboration. These issues manifest themselves through fragmented communication, isolated work practices, and imbalanced decision-making, contributing to social debt—organizational inefficiencies that hinder productivity. One notable example is the Prima Donna Effect, where dominant individuals disregard others' input, reducing collaboration and team morale~\cite{tamburri2015social,almarimi2021_csdetector}.

The relevance of community smells is reinforced by the concept of socio-technical congruence, which highlights the interdependence between social and technical structures in software projects~\cite{tamburri2015social}. Studies suggest that unresolved social issues can influence software quality, as shown by \citeauthor{palombaTechnicalAspectsHow2021}~\cite{palombaTechnicalAspectsHow2021}, who found that community smells are linked to persistent code smells and technical debt. Developers often attribute software quality issues to communication breakdowns and coordination challenges within teams. 
Several factors influence the emergence of community smells~\cite{cataldo2006identification,Catolino2019GenderDiversity,lambiaseGoodFencesMake2022,annunziata2024empirical}. Gender diversity, for instance, has been associated with reduced occurrences of these issues, as diverse teams tend to foster more inclusive and communicative environments~\cite{Catolino2019GenderDiversity}. Similarly, cultural and geographical diversity can introduce both benefits and challenges: while it encourages innovation, it may also lead to coordination difficulties in distributed teams~\cite{lambiaseGoodFencesMake2022}. Additionally, leadership roles such as architects and team leaders often become knowledge bottlenecks, where critical expertise is concentrated in a few individuals, leading to inefficiencies if knowledge is not effectively shared~\cite{cataldo2006identification}.

Mitigation strategies have been explored to reduce the impact of community smells. \citeauthor{catolino2021understanding}~\cite{catolino2021understanding} identified approaches such as improved communication, structured collaboration, and inclusive decision-making to address these socio-technical challenges. 

While community smells have been extensively studied in traditional software development, their role in ML-enabled systems remains underexplored. Given the heterogeneous nature of ML communities—often composed of data scientists, software engineers, and domain experts—the risk of socio-technical misalignment is particularly high. Initial investigations suggest that community smells may manifest differently in ML projects. \citeauthor{annunziata2024empirical}~\cite{annunziata2024empirical} found that in open-source projects using Python, a language dominant in ML, the Prima Donna Effect is more prevalent, indicating potential social challenges in ML development. Expanding on this, \citeauthor{annunziata2025Uncovering}~\cite{annunziata2025Uncovering} identified correlations between community smells and identified socio-technical anti-patterns in the communities of ML-enabled systems, positioning these patterns as both causes and consequences of organizational inefficiencies.

These findings suggest that community smells could serve as indicators for assessing the social health of teams of ML-enabled systems, offering a new lens to study and mitigate challenges unique to this context. However, the extent to which heterogeneous teams influence the persistence of community smells in ML-enabled projects remains largely unexplored. Addressing this gap is crucial to understanding the social dynamics of ML-enabled system development and improving team collaboration, system maintainability, and overall software quality.

\stesummarybox{\faOutdent\ Research Gap}{Social aspects in software development often influence technical components, ultimately affecting system quality. In ML-enabled systems, the interdisciplinary nature of heterogeneous teams amplifies these challenges. While crucial, the study of socio-technical issues—such as measurable patterns like 'community smells'—remains partially underexplored compared to technical challenges. This gap underscores the need to better understand their impact on ML teams and system quality. Investigating the prevalence of community smells could initially provide valuable insights to mitigate these challenges and guide future research on possibly improving team performance in ML environments.}

%% file: Sections/3.rm.tex
\section{Research Method}
\label{sec:rm}

The \textit{goal} of the study was to understand and examine the socio-technical well-being of communities developing ML-enabled systems using community smells. The \textit{purpose} gains knowledge about the role of social anti-patterns in such communities to inform new research trends and provide useful knowledge to practitioners.
The study is primarily conducted from the perspective of project managers, who seek to identify and address social anti-patterns within their heterogeneous ML-enabled teams to prevent potential negative impacts on technical outcomes. A secondary perspective is that of researchers, offering them insights into the prevalence and consequences of community smells in ML projects and laying the groundwork for further investigations.
To achieve the above-mentioned \textit{objective}, we defined a set of research questions that guided the research process.
First, we wanted to characterize the extent to which socio-technical anti-patterns—represented using community smells—are present in ML-enabled open-source communities, i.e., their prevalence.

Thus, we formulated the following research question:

\steattentionbox{\faSearch\ \textbf{RQ\textsubscript{1}} What is the prevalence of Community Smells in ML-enabled projects?}

The initial analysis offered a broad overview of the prevalence of community smells across these projects. To build on this, we are conducting a more detailed time-slice analysis to achieve two primary objectives: (1) identify which community smells are most prevalent over time, and (2) develop a deeper understanding of these social issues through a more comprehensive knowledge base over time. 
To guide this investigation, we have formulated the following research question:

\steattentionbox{\faSearch\ \textbf{RQ\textsubscript{2}} How does the Prevalence of Community Smells in ML-enabled projects vary over time?}

Then, we focus on analyzing the correlations between different smells. With this analysis, we wanted to deepen our understanding of community smells by shedding light on how these issues interact with each other and shape team dynamics. More specifically, the analysis highlights correlations among the smells that suggest that the presence of one smell may indicate the likelihood of others occurring more or less frequently. In other words, the presence of a positive correlation between two smells informs us that when one occurs, there is a high probability that the other smell will also occur. In contrast, the presence of a negative correlation suggests that, upon the occurrence of one smell, the probability of the other smell occurring is lower than normal. This insight can be valuable to practitioners, enabling them to anticipate and address related smells once one is identified. 

Thus, we formulated the following research question:

\steattentionbox{\faSearch\ \textbf{RQ\textsubscript{3}} What are the relationships between different community smells in the ML-enabled projects?}

\begin{figure*}[h]
    \centering
    \includegraphics[width=1\linewidth]{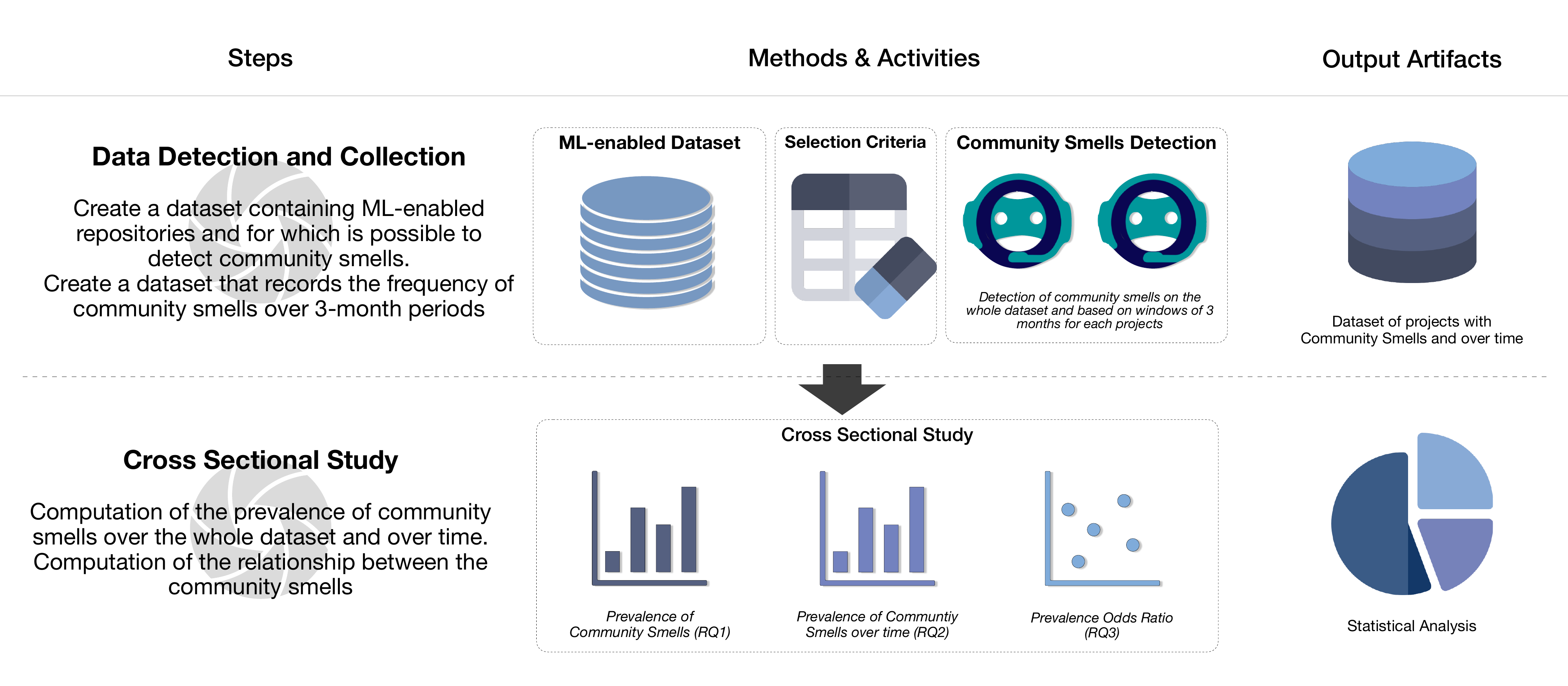}
    \caption{Research Method}
    \label{fig:rm}
\end{figure*}

To answer our research questions, we conducted a cross-sectional study in combination with a mining study on GitHub. A \textit{Cross-sectional} studies provide a snapshot of a population at a specific point in time, capturing the prevalence of a condition or phenomenon and relevant exposures by collecting data simultaneously from all participants~\cite{wang2020cross}. Specifically, we analyzed 188 open-source ML-enabled projects belonging to the NICHE dataset~\cite{widyasari2023niche}. As for \textit{RQ\textsubscript{1}}, we calculated the prevalence of community smells within the 188 \textsc{GitHub} ML-enabled open-source projects. For \textit{RQ\textsubscript{2}}, we performed a longitudinal analysis, dividing the timeline into 3-month intervals to examine the evolution of these community smells over time. Finally, for \textit{RQ\textsubscript{3}} we conducted a cross-sectional study using the \textit{Prevalence Odds Ratio} (POR) to identify potential correlations between different types of community smells.
Figure~\ref{fig:rm} presents an overview of our study. In particular, reports the steps, the methods, and activities to conduct it.

To conduct our analysis, we follow the \emph{Empirical Standards} by \citeauthor{ralph2020empirical}~\cite{ralph2020empirical}. Specifically, we employed the \textsl{``Repository Mining''}, the \textsl{``Longitudinal''}, the \textsl{``Engineering Research''} and the \textsl{``General Standard''} definitions and guidelines.\footnote{Empirical Standards: \url{https://www2.sigsoft.org/EmpiricalStandards/docs/standards}}


\subsection{Context of the Study and Dataset Creation}

To investigate socio-technical issues in the context of ML-enabled systems, we began by creating a dataset containing community smells on ML-enabled open-source projects. As a foundation for this dataset, we used the \textsc{NICHE} dataset developed by \citeauthor{widyasari2023niche}~\cite{widyasari2023niche}. The \textsc{NICHE} dataset, validated in 2022, consists of 572 open-source machine learning (ML) projects implemented in Python and labeled manually. Of these, 441 are classified as “engineered” because they clearly adhere to good software engineering practices, and 131 are classified as “non-engineered” because they do not meet engineering standards~\cite{widyasari2023niche}.
The dataset obtained is available in the online appendix~\cite{online_appendix}.


\subsubsection*{\textbf{Selection Criteria}}
\label{lab:secCriteria}
In this section, we discuss the selection criteria applied to obtain our dataset. Our criteria are based on the objective of the study and rely on criteria adopted in similar previous studies on the matter~\cite{catolino2021understanding}.

\paragraph{Engineered Project:}
The dataset \textsc{NICHE}~\cite{widyasari2023niche} was filtered to exclude non-engineered projects. 
This decision was driven by the objective of studying human factors and behavioral aspects in environments involving data scientists and software engineers~\cite{Busquim2024interactionSEDS,ozkaya2020really}. ML-enabled projects inherently feature an ML component, insinuating the presence of data scientists into the development team, while engineered projects suggest the inclusion of software engineers. This combination aligns with our goal of analyzing socio-technical interactions specific to ML-enabled development.

\paragraph{Age of the Project:} Since the second research question focuses on analyzing the evolution of community smells over time, project age was used as a key selection criterion. Projects were categorized into 3 groups based on their lifespan: Young (<24 months), Established (24–32 months), and Mature (>32 months). Most of the selected projects fell into the Established (152) and Mature groups(207).

We analyzed projects in 3-month windows, following the approach used in previous studies~\cite{catolino2021understanding}. For consistency, each age group was analyzed over \textit{a number of windows corresponding to the youngest project in that category}: 6 months (2 windows) for Young, 24 months (8 windows) for Established, and 32 months (10 windows) for Mature. This ensured comparability across projects within each age group.
Due to the low number of projects and time windows in the Young category (82), we decided to exclude it, focusing only on Established and Mature projects. Furthermore, because the Established category had fewer projects than the Mature category, we merged these two groups into a single category. This consolidation allowed for a consistent analysis timeframe of up to 8 time windows (equivalent to 2 years) for all selected projects, ensuring sufficient data while maintaining analytical robustness.

\paragraph{Size of the Team:} Following conventions in the literature~\cite{tamburri2015open-source,catolino2021understanding}, projects were categorized based on the number of contributors: \textit{Small} (Fewer than 10 contributors), \textit{Medium} (10–50 contributors), \textit{Large} (50–150 contributors), and \textit{Very Large} (More than 150 contributors).
The dataset had a low number of small projects, with very small teams, less than five or even one team member. A low number of team members is insufficient to analyze community smells on these projects, so the projects identified as small were excluded from the dataset. 
Our focus shifted to medium, large, and very large projects to ensure meaningful socio-technical insights, obtaining our dataset with 188 projects.

\subsection{Community Smells Detection}

We began with an initial dataset of 572 ML-enabled projects sourced from the validated \textsc{NICHE} dataset~\cite{widyasari2023niche} (2022). Following the selection criteria explained, we refined the pool to 188 projects. On those projects we used \textsc{CADOCS}~\cite{voria2022_CADOCS} to detect community smells.
\textsc{CADOCS} is a client-server conversational agent integrated into Slack, designed to extend and facilitate the use of the community smell detection tool `\textsc{CSDetector}'.\footnote{CSDetector using classification metrics, achieving AUC values between 0.89 and 0.96 (average 0.93) and F1 scores ranging from 0.77 to 0.89} 
The goal of \textsc{CADOCS} is to provide practitioners with more social problem analysis support from a GitHub repository, leveraging a graphical interface and client-server communication.
\textsc{CADOCS}, such as \textsc{CSDetector}~\cite{almarimi2021_csdetector}, is able to detect 10 different Community Smells, illustrated in Table~\ref{table:community_smells}.

Following the initial detection, we conduct a longitudinal analysis by segmenting the lifespan of each project into eight consecutive 3-month windows, covering a total period of 2 years. Within each time window, we apply \textsc{CADOCS} to generate individual datasets, enabling a detailed temporal examination of community smells and their evolution over time.
 
The final dataset obtained starts from the validated data from the \textsc{NICHE} dataset~\cite{widyasari2023niche}, filtered according to the selection criteria explained, and expanded with the community smells for the projects that \textsc{CADOCS}~\cite{voria2022_CADOCS} was able to detect in December 2024. Through this process, we created our dataset with 188 projects. Detailed reports on this dataset, including the time-windowed datasets (one for each project) are available in our online appendix~\cite{online_appendix}.
 
\begin{table*}[h]
 \small
 \centering 
 \caption{Community Smells.}
 
 \rowcolors{1}{graytable}{white}
 
 \resizebox{\linewidth}{!}{
  \begin{tabular}{|p{0.18\linewidth}p{0.02\linewidth}p{0.78\linewidth}|}
  \rowcolor{black}
  \textcolor{white}{Community Smells} &\textcolor{white}{AC} &\textcolor{white}{Definition~\cite{almarimi2021_csdetector}} \\
  \textbf{Organizational Skirmish } & \textbf{OS}& Team members with different levels of competence lead to a drop in productivity, impacting time and costs.
  \\
  \textbf{Black Cloud Effect } & \textbf{BC} & A lack of structured communications or cooperative governance can lead to information overload.
  \\
  \textbf{Radio Silence} & \textbf{RS}& The use of formal communication between multiple sub-communities penalizes flexibility and causes loss of time.
  \\
  \textbf{Prima Donnas Effect } & \textbf{PDE}& Team members that expose condescending behavior, superiority, constant disagreement, and uncooperativeness.
  \\
  \textbf{Sharing Villainy }& \textbf{SV}& Failure to exchange information can lead team members to share incorrect, obsolete, and unconfirmed information.
  \\
  \textbf{Organizational Silo Effect} & \textbf{OSE}& Siloed community that do not communicate with each other except through one or two members.
  \\
  \textbf{Solution Defiance} & \textbf{SD}& Having a community with different backgrounds of experience and culture can lead to subgroups with conflicting opinions.
  \\
  \textbf{Truck Factor Smell }& \textbf{TFS}& Developer turnover may cause a significant loss of knowledge as it is concentrated in a minority of developers.
  \\
  \textbf{Unhealthy Interaction} & \textbf{UI}& Slow, light, and short conversations and discussions caused by long delays in stakeholder communication.
  \\
  \textbf{Toxic Communication} &\textbf{TC}& Developers may negatively interact with their colleagues, leading to frustration, stress, and project abandonment.
  \\
  \hline
  \end{tabular}
  }
 \label{table:community_smells}
\end{table*}


\subsection{Data Analysis}

In this section, we provide a general overview of cross-sectional studies, explaining their current use in the context of software engineering, and then detail how we applied them in our specific study to address our research questions.

\subsubsection{Overview on Cross-Sectional Studies}

The field of software engineering research has experienced a notable surge in studies leveraging mining software repositories (MSRs), fueled by the growing popularity of online code hosting platforms like \textsc{GitHub}. This trend has prompted the development of best practices and guidelines to mitigate common pitfalls and enhance the reliability of these studies and their findings~\cite{kalliamvakou2014promises}.

Despite their utility and relative simplicity, MSR studies face a key limitation: they are unable to establish causal relationships for the phenomena they observe. To overcome this shortcoming, \citeauthor{saarimaki2020cohort}~\cite{saarimaki2020cohort} have advocated for the use of observational methods, particularly cohort studies, which provide a more robust foundation for generating high-quality scientific evidence.
For this reason, the adoption of a cross-sectional study is a good starting point for addressing this investigation of mining repositories in the context of community smells in ML-enabled systems. 
Since ML-enabled devices are considered an emerging discipline, the social-technical aspects are still not well explored.

Cross-sectional studies aim to capture and describe the characteristics of a population at a specific point in time. They offer a snapshot of the prevalence of a particular condition or phenomenon and the distribution of relevant exposures within the population~\cite{wang2020cross}. Unlike other observational study designs, cross-sectional studies collect data simultaneously from all participants. 
The \textit{Prevalence} represents the proportion of individuals in a population who exhibit a specific condition or characteristic at a given moment~\cite{wang2020cross}; in our study, the presence of community smells. Moreover, the \textit{Prevalence Odds Ratio} (\textbf{POR}) is often employed to assess the strength of the association between exposure and a condition. The POR compares the odds of exposure among those with the condition to the odds of exposure among those without it, offering a valuable measure of the relationship in the sample~\cite{wang2020cross}.

Cross-sectional studies provide a cost-effective and pragmatic approach for initiating the exploration of socio-technical issues in ML-enabled systems. In this context, such problems are represented by community smells and their interrelationships, given by their heterogeneity, such as they are composed by data scientists and software engineers~\cite{Busquim2024interactionSEDS}. Research on the socio-technical dimensions of ML-enabled systems is a relatively new and unexplored field~\cite{ML-EnabledSystemsCCDE}. Cross-sectional studies can deliver critical insights into the current state of this domain, revealing conditions and potential associations. These studies are particularly effective for generating hypotheses and identifying possible connections. Although they do have limitations, cross-sectional studies serve as a practical and foundational step in the advancement of research in this area.

\vspace{5mm}
\subsubsection{\textbf{Answering} \textbf{RQ\textsubscript{1}}}

To answer to \textit{RQ\textsubscript{1}}, we calculate the \textit{Prevalence} of each detected community smell as the ratio of repositories affected by a specific smell to the total number of repositories analyzed~\cite{wang2020cross}. Formally, it is defined as:

\begin{equation}\label{eq_prelavelence_formula}
P(X) = \frac{\text{Number of repositories affected by smell X}}{\text{Number of all Repository}}
\end{equation}
where P(X) is the prevalence of the smell X in the dataset.

\subsubsection{\textbf{Answering} \textbf{RQ\textsubscript{2}}}
To answer to \textit{RQ\textsubscript{2}}, we calculate the \textit{Prevalence} for each community smell over time; we analyze the projects in the created dataset using \textsc{CADOCS} with a 3-month temporal granularity, segmenting the data into distinct time windows. These windows span from the first commit of each project up to 2 years of its lifecycle, as explained in Section~\ref{lab:secCriteria}. The 3-month granularity was chosen based on socio-technical metrics, which consider this interval sufficient to capture potential turnover and provide an optimal window for studies of this nature~\cite{huang2021predicting,palombaPredictingEmergenceCommunity2021,catolino2021understanding}. Once we obtained the detection of the smells in the time windows for each project, we calculated the prevalence of each community smells for each time window.

\subsubsection{\textbf{Answering} \textbf{RQ\textsubscript{3}}}
To answer to \textit{RQ\textsubscript{3}}, we calculate the \textit{Prevalence Odds Ratio (\textbf{POR})}, which measures the association between the presence or absence of a specific community smell and the presence or absence of another community smell~\cite{wang2020cross}. This metric helps evaluate whether certain community smells are more likely to co-occur within ML-enabled projects, providing insights into their interrelationships. It is calculated as: 
\begin{equation}\label{eq_POR_formula}
POR=\frac{AD}{BC}
\end{equation}

To illustrate this with an example, consider the calculation of POR between Organizational Silo Effect (OSE) and Prima Donna Effect (PDE). In this context, \textit{A} represents the number of cases where both community smells, OSE and PDE, are present. \textit{B} denotes instances where OSE is present, but PDE is absent. \textit{C} refers to cases where OSE is absent, but PDE is present. Finally, \textit{D} represents instances where both OSE and PDE are absent.

Regarding the interpretation of the results, when the value of POR is less than 1, it indicates that there is a negative correlation, in our case, that the presence of the second smell decreases the likelihood of occurrence of the first smell. In the case of a POR of 1 or higher, it indicates a strongly positive association in which the presence of the second smell increases the presence of the first one. 
For example, a POR value of \textit{2} indicates that the community smell OSE is twice as likely to occur in communities where the community smell PDE is present compared to those where it is absent. Conversely, a POR value of \textit{0.5} suggests that OSE is half as likely to occur in communities with the presence of PDE, implying that PDE may have a mitigating effect.

%% file: Sections/4.results.tex
\section{Analysis of Results}
\label{sec:results}

This section presents the main findings of our study, detailing the results of the cross-sectional analysis and providing insights into the observed patterns and trends.

\subsection{RQ\textsubscript{1}—Prevalence of Community Smells}

\begin{figure}
  \centering
  \includegraphics[width=1\linewidth]{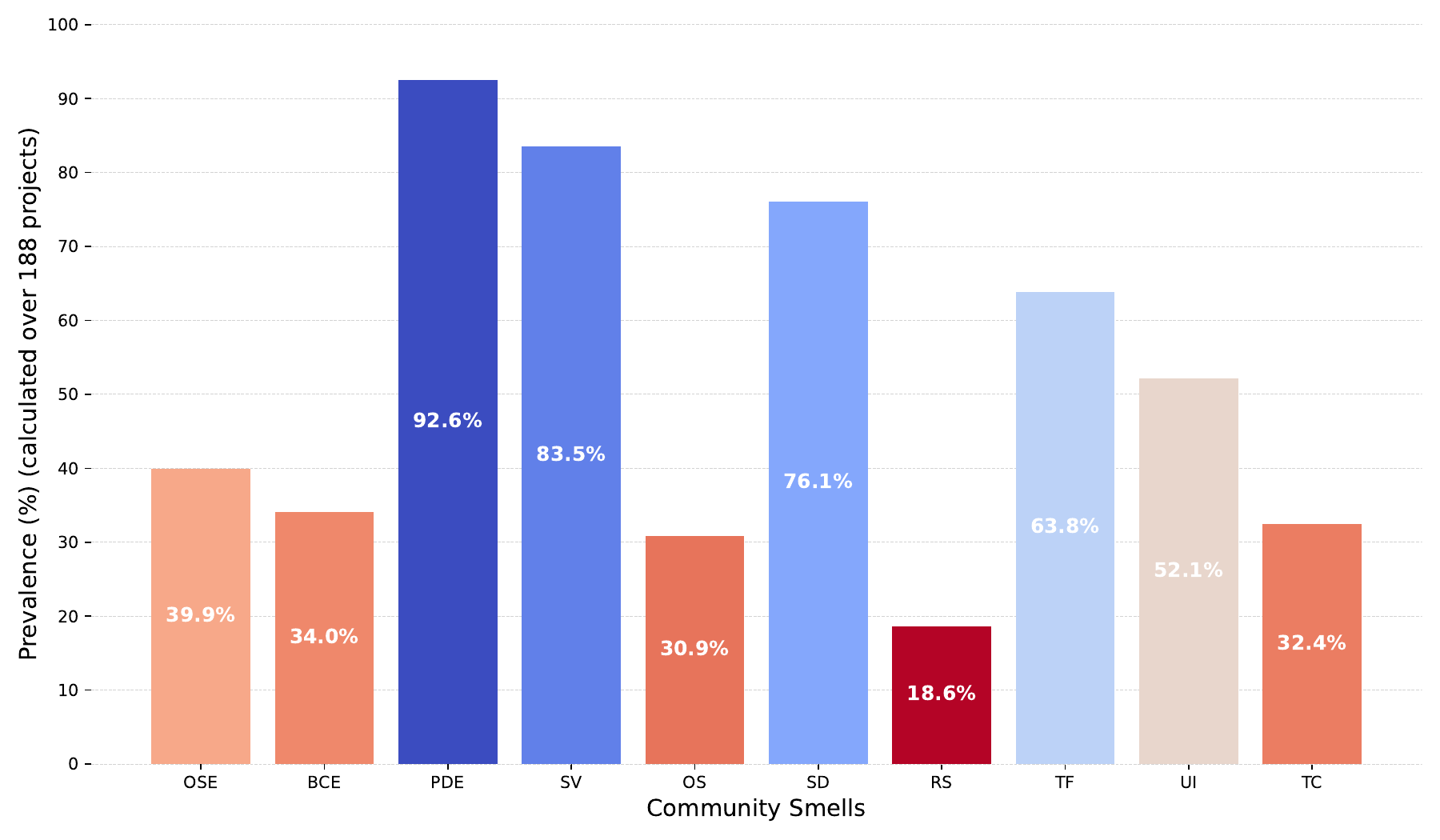}
  \caption{Prevalence of Community Smells in ML 
Projects}
  \label{fig:prevalence}
\end{figure}

To answer to \textit{RQ\textsubscript{1}} we calculated the prevalence of each community smell across the 188 \textsc{GitHub} ML-enabled open-source projects of our dataset. Figure~\ref{fig:prevalence} shows the distribution of community smell prevalence within the analyzed projects.

Examining the prevalence distribution, we observe that half of the community smells have a prevalence greater than 50\%. 
In particular, Prima Donna Effect (PDE) stands out with a prevalence of 92.6\% indicating its presence in nearly all the analyzed projects.
The second most prevalent smell is Sharing Villainy (SV), with a prevalence of 83.5\%. 
Another noteworthy smell is Solution Defiance (SD), which registers a significant 76.1\%. 
Conversely, some community smells exhibit low prevalence, indicating their limited presence within the analyzed projects. For example, Radio Silence (RS)shows a prevalence of 18.6\%. 
Similarly, Organizational Skirmish (OS) appears in only 30.9\% of the projects, indicating its rarity compared to the more widespread community smells.

\subsection{RQ\textsubscript{2}—Prevalence in Time Windows}

\begin{figure*}
  \centering
  \includegraphics[width=0.95\linewidth]{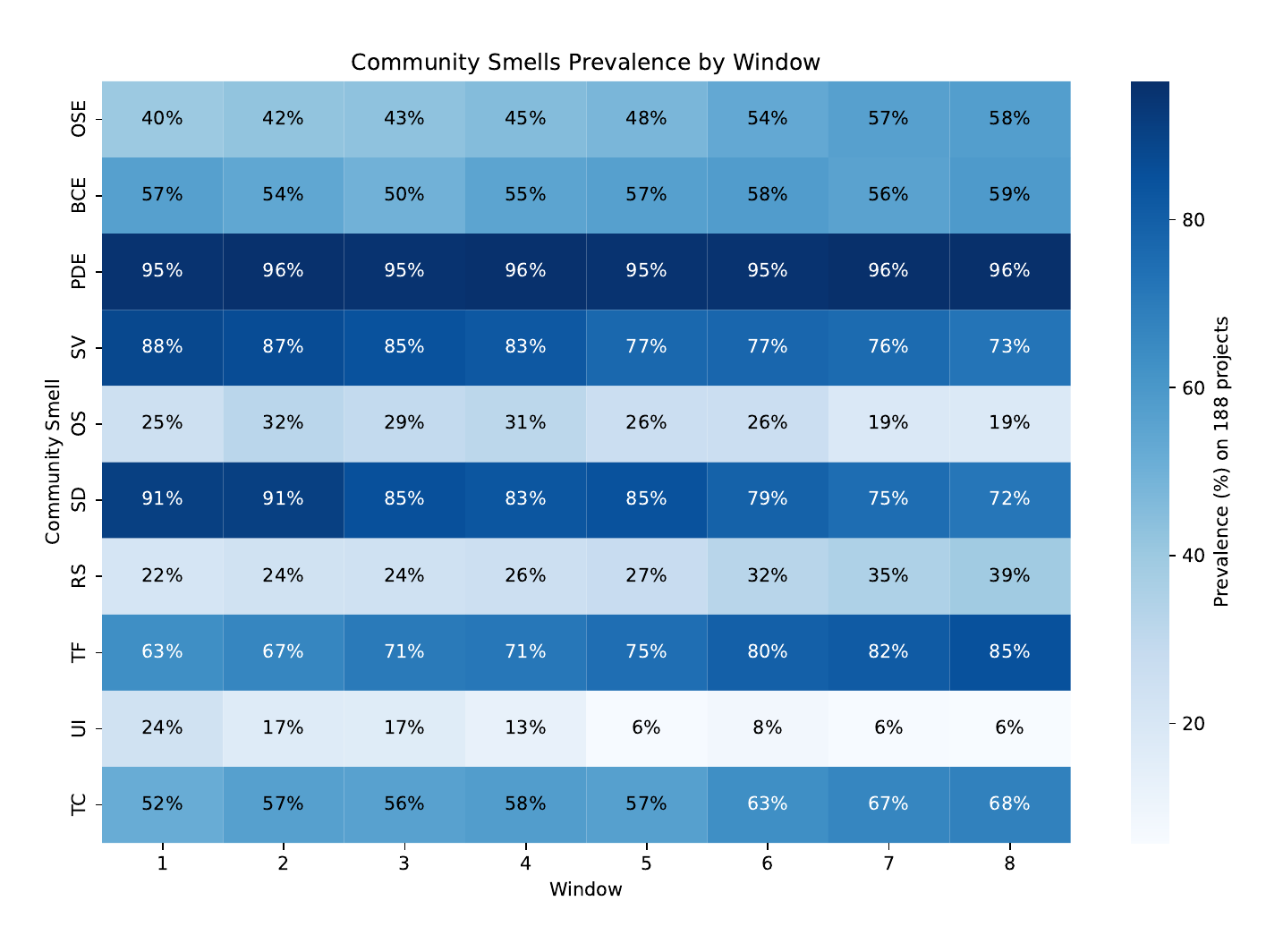}
  \caption{Prevalence of Community Smells in ML 
Projects in Windows of 3 Months}
  \label{fig:prevalence3months}
\end{figure*}

Figure~\ref{fig:prevalence3months} shows a heatmap where each row represents a community smell, and each column corresponds to the time window analyzed. Each time window represents 3 months and ranges from the beginning of the project to 2 years. Each cell represents the prevalence of a community smell within a specific time window.

From an initial analysis, we note that the heatmap reflects the overall prevalence trends of the projects, where Prima Donna Effect (PDE), Sharing Villainy (SV), and Solution Defiance (SD) remain the most frequent smells, while Organizational Skirmish (OS) and Radio Silence (RS) continue to exhibit low prevalence. 

Analyzing them in detail, we noticed interesting results.

PDE emerges as the most persistent smell, with a prevalence ranging between 95\% and 96\%, higher than the 92.6\% reported in the overall project analysis. 
However, for other smells the trend is less stable. Sharing Villainy (SV), Solution Defiance (SD), and Unhealty Interaction (UI) exhibit a steady decrease over time. More in detail, we see that SV started with a prevalence of 88\% in the first quarter and dropped to 73\% after 2 years, SD decreases from 91\% to 72\%, and UI shows a decline from 24\% to just 6\%.
A different discussion can be made for the community smell Organizational Skirmish (OS). It started at 25\% in the first quarter, peaked at 32\% in the second, and then declined steadily to 19\% in the last quarter analyzed.
In contrast, Organizational Silo Effect (OSE), Radio Silence (RS), Truck Factor (TF), and Toxic Communication (TC), exhibit a continuous increase over time: OSE rised from 40\% to 58\%, RS increased from 22\% to 39\%, TF climbed from 63\% to 85\%, and TC growed from 52\% to 68\%. 
A different pattern emerges for Black Cloud Effect (BCE), which started at 57\%, decreased in the third quarter to 50\%, and then increased to 59\%. 

\subsection{RQ\textsubscript{3}—Prevalence Odds Ratio Between Community Smells}

\begin{figure}
  \centering
  \includegraphics[width=1\linewidth]{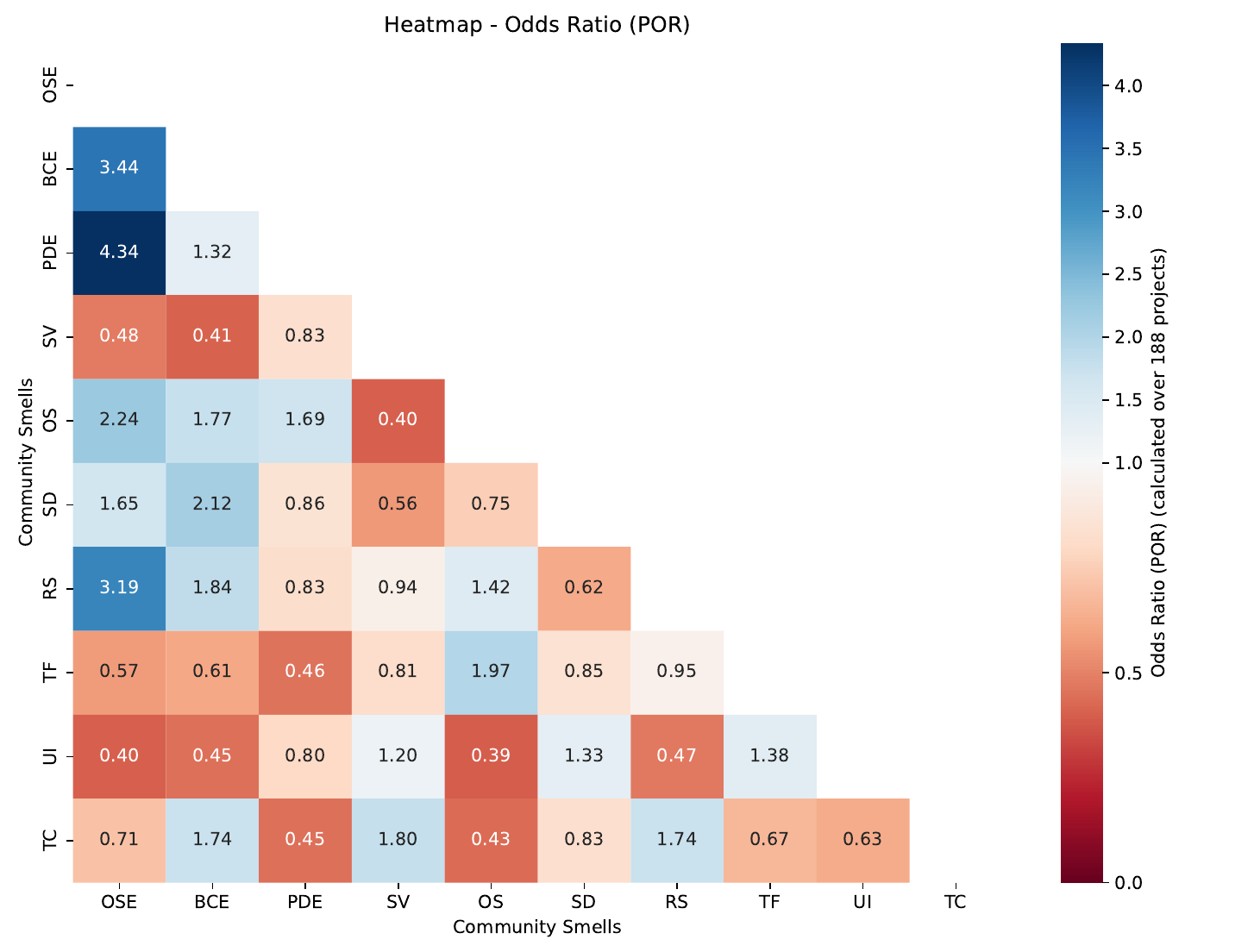}
  \caption{Heat map of Prevalence Odds Ratio}
  \label{fig:oddsratio}
\end{figure}

Figure \ref{fig:oddsratio} presents a heatmap displaying the Prevalence Odds Ratio (POR) values among the various community smells analyzed in the 188 projects.
The first notable insight is that most correlations between community smells are negative.
However, although inferior, the positive correlations tend to be more intense. 
An example is the correlation between Prima Donna Effect (PDE) and Organizational Silo Effect (OSE), which stands out with a POR of 4.34. This strong association suggests that it is very probable to find the community smell PDE when the community smell OSE occurs. 
Similarly, Black Cloud Effect (BC) and Radio Silence (RS) also show a high correlation with OSE, with POR values of 3.44 and 3.19, respectively.
Other notable positive correlations, although more moderate, are between Organizational Skirmish (OS) and OSE, with a POR of 2.24, between Solution Defiance (SD) and BCE, with a POR of 2.12, and between Truck Factor (TC) and Organizational Skirmish (OS) with a POR of 1.97. 
Other moderate positive correlations are observed between PDE and ECB (POR 1.32), OS and ECB (POR 1.77), OS and PDE (POR 1.69), SD and OSE (POR 1.65), RS and ECB (POR 1.84), RS and OS (POR 1.42), Unhealty Interaction (UI) and Sharing Villainy (SV) (POR 1.20), UI and Solution Defisnce (SD) (POR 1.33), UI and TF (POR 1.38), Toxic Communication (TC) and BCE (POR 1.74), TC and SV (POR 1.80), and TC and RS (POR 1.74).

Correlations with POR values below 1 indicate a negative relationship. For instance, UI and OS show a strong negative correlation (POR 0.39), meaning that the presence of UI reduces the likelihood of OS occurring. Other high negative correlations are between OS and SV smells (POR 0.40), IU and OSE (POR 0.40), SV and ECB (POR 0.41), IU and ECB (POR 0.45), TF and PDE (POR 0.46), TC and OS (POR 0.43), IU and RS (POR 0.47), and TC and PDE (POR 0.45). 

\stesummarybox{\faBraille \hspace{0.05cm} Key Results}
{ 
  \begin{itemize}[noitemsep,topsep=0pt,leftmargin=*]
    \item PDE and SV are the community smells most present in ML-enabled projects (92.6\% and 83.5\%).
    \item PDE always remains constant, while the other smells steadily increase or decrease over time.
    \item Most correlations between community smells are negative, but the positive correlations tend to be stronger.
  \end{itemize}
}

%% file: Sections/5.discussion-implication.tex
\section{Discussions and Implications}
\label{sec:dis_imp}

In this section, we discuss the results of the study and focus on their implications, both for researchers and practitioners.

\subsection{Prevalence of Community Smells}

The results highlight the common occurrence of community smells in ML-enabled projects, with certain types being especially pervasive.
Prima Donna Effect (PDE) stands out with a prevalence of 92.6\%, indicating that dominant, non-collaborative behaviors among individual developers are a persistent issue in the analyzed projects. This suggests a critical socio-technical challenge, where influential team members impose their decisions, potentially stifling collaboration and innovation. The high prevalence of this behavior in ML-enabled projects might suggest that it may often be adopted by Data Scientists, as already highlighted by literature~\cite{annunziata2025Uncovering,SocioTechnicalAntiPattern,ozkaya2020really}. 
Prior studies ~\cite{annunziata2025Uncovering} associated PDE with the tendency of data scientists to assert expertise, which can lead to conflicts within team members and frustration among project managers, who struggle to manage these dynamics effectively~\cite{annunziata2025Uncovering,SocioTechnicalAntiPattern}.
Sharing Villainy (SV), with a prevalence of 83.5\%, highlights the widespread lack of adequate information sharing. This often results in the spread of incorrect or outdated knowledge, which can negatively impact project quality and decision-making. 
Similarly, Solution Defiance (SD), occurring in 76.1\% of projects, indicates  frequent subgroups conflicts, resulting in decision-making bottlenecks~\cite{Busquim2024interactionSEDS} and resistance to established best practices.
The high prevalence of PDE, SV, and SD reflects the inherent challenge of aligning the workflows and goals of data scientists and software engineers. This misalignment fosters power imbalances and conflicting goals, leading to community breakdown~\cite{annunziata2025Uncovering,SocioTechnicalAntiPattern,ozkaya2020really}.

Conversely, lower prevalence rates of Radio Silence (RS) (18.6\%) indicate that explicit breakdowns in communication or role-based conflicts are less pervasive but still significant. 
Organizational Skirmish (OS) (30.9\%) also presents a low prevalence. In the ML-enabled context, this community smell can arise when data scientists and software engineers lack the skills to adequately understand one another. Their distinct backgrounds can hinder effective collaboration and impact project outcomes ~\cite{SocioTechnicalAntiPattern,ozkaya2020really}. However, the relatively lower occurrence of OS and RS suggests that differences in knowledge and expertise may not directly contribute to social issues or technical challenges within the project~\cite{annunziata2025Uncovering}.
The relatively lower occurrence of these smells may reflect the presence of structured workflows, although informal knowledge sharing gaps persist.

Moreover, comparing our results with those in the literature~\cite{tamburri2015open-source,communitysmellsSLR}, we note how the distribution of smells varies in the context of ML-enabled systems. An example are Organizational Silo (OSE) and Radio Silence (RS), which are very present in traditional systems, but with a low prevalence in ML-enabled systems. 
Meanwhile, there is an alignment with the literature about the low presence of Black Cloud smell~\cite{communitysmellsSLR,tamburri2015social}.

\begin{itemize}[leftmargin=0.5cm]
     \item[\faBook] \textbf{Implications for research:} Further investigation is needed to understand how the distinct goals of data scientists and software engineers contribute to the emergence of community smells PDE and SV and mitigate their prevalence.

     \item[\faBriefcase] \textbf{Implications for practice:} Encouraging structured collaboration between data scientists and software engineers can help reduce community smells. For example, rotating responsibilities between data scientists and software engineers or adding intermediate figures between the two roles could reduce the occurrence of such smells in teams~\cite{SocioTechnicalAntiPattern}.
\end{itemize}

\subsection{Evolution of Community Smells Over Time}

The longitudinal analysis reveals nuanced trends in the evolution of community smells. 
While PDE maintains a consistently high prevalence over time (between 95\% and 96\%), indicating persistent dominance behaviors, SV and SD show gradual declines. 
This result indicates a consistent pattern among one or more team members exhibiting superiority and limited collaboration with the rest of the team. Literature suggests that such behaviors are often attributed to data scientists, who tend to prioritize their individual goals over the project objectives~\cite{SocioTechnicalAntiPattern,annunziata2025Uncovering}. A persistent and significant presence of these smells highlights the recurrence of such behavior throughout the software development process~\cite{annunziata2025Uncovering}.

SV drops from 88\% to 73\% over 2 years, while SD decreases from 91\% to 72\%. This suggests that initial conflicts and isolated knowledge practices may reduce as teams establish shared lines of conduct and governance structures~\cite{SocioTechnicalAntiPattern}. 
However, smells such as Organizational Silo Effect (OSE) and Toxic Communication (TC) increase over time, with OSE rising from 40\% to 58\% and TC from 52\% to 68\%. 
These results suggest that when projects become larger, it may be the case that split subgroups between data scientists and software engineers come into existence; these subgroups may operate independently, leading to fragmented communication and may emerge a separation of roles and responsibilities~\cite{annunziata2025Uncovering}. These problems often result in less efficient collaboration and hinder the overall success of software development projects. 

\begin{itemize}[leftmargin=0.5cm]
     \item[\faBook] \textbf{Implications for research:} Conducting more in-depth longitudinal studies and considering causal factors of such community smells, could provide a more complete and detailed picture of the presence and evolution of such smells in ML-enabled development teams.

     \item[\faBriefcase] \textbf{Implications for practice:} The findings suggest an increase in OSE and TC smells over time; to counteract this, practitioners could adopt regular team reviews and promote open communication channels~\cite{SocioTechnicalAntiPattern,annunziata2025Uncovering}.
\end{itemize}

\subsection{Correlation Between Community Smells}

The cross-sectional analysis identifies different positive correlations between community smells. 
Prima Donna Effect (PDE) correlates strongly with Organizational Silo Effect (OSE) (Prevalence Odds Ratio, POR = 4.34), suggesting that teams dominated by a lead data scientist or software engineer are more prone to isolation and lack of cross-group communication. 
This result suggests a strong coexistence of the two community smells. It might be reasonable to think that the adoption of Prima Donna Effect behavior, adopted mostly by data scientists when they aim for their individual goals, tends to estrange other team members, causing the team to ``break'' into uncooperative sub-teams and isolated workflows driven by data scientists prioritize experimentation or software engineers who emphasize delivery timelines~\cite{annunziata2025Uncovering}.

Similarly, the Black Cloud Effect (BCE) correlates with Radio Silence (RS) (POR = 3.19), highlighting scenarios where insufficient governance leads to information bottlenecks.
Negative correlations, such as between Unhealthy Interaction (UI) and Organizational Skirmish (OS) (POR = 0.39), suggest that while direct conflict may occur, it can deter large-scale division by forcing collaboration. 
BCE and RS often result from gaps in communication channels, where the absence of structured leadership contributes to limited information sharing.

\begin{itemize}[leftmargin=0.5cm]
     \item[\faBook] \textbf{Implications for research:} More research is needed to uncover more details about the rationale behind the behaviors of data scientists and software engineers that lead to the emergence of community smells. This may lay the basis for further work related to correlation or association rules between different community smells occurring in the ML-enabled context.

      \item[\faBriefcase] \textbf{Implications for practice:} Professionals should apply strategies to mitigate social issues; \eg apply pair programming or introduce team members who can communicate and actively collaborate with both developers role~\cite{annunziata2025Uncovering,SocioTechnicalAntiPattern}.
\end{itemize}

\subsection{Differences in Prevalence of Community Smells between RQ1 and RQ2}
Interestingly, a discrepancy emerges when comparing the prevalence of some community smells between the cross-sectional and longitudinal analyses, i.e., the prevalence of the whole projects and the prevalence in windows of 3 months from birth to 2 years of life for each project. 
Prima Donna Effect (PDE) reported at 92.6\% in the initial analysis (RQ1), increasing to 95-96\% in the longitudinal study (RQ2). This slight increase may indicate that the dominance of some developers, often Data Scientists or Principal Software Engineers, intensifies as projects evolve and key collaborators consolidate their roles. In contrast, Unhealthy Interaction (UI) shows a reverse trend. Initially, UI showed a prevalence of 52\% in RQ1, but over time, it fluctuated between 24\% and 6\% in the longitudinal analysis. This decline might suggest that, while communication problems are evident early in the project life cycle, teams gradually implement mechanisms to reduce their frequency, such as regular stand-ups or communication protocols.
This discrepancy in prevalence poses an excellent baseline question to entice longitudinal studies, with the goal of studying in detail the evolution and motivations behind community smells in the ML-enabled context.


\stesummarybox{\faAngleDoubleRight \hspace{0.05cm} Take Away Results}
{ 
    \begin{itemize}[noitemsep,topsep=0pt,leftmargin=*]
        \item The constant and high prevalence of the smell Prima Donna Effect suggests misaligned workflows between data scientists and software engineers, which hinders team collaboration and productivity.
        \item Community smells are not static, they can decrease (e.g., SV, SD, and UI), or increase (e.g., OSE, RS, TF, and TC).
        \item Most correlations between community smells are negative, but there are cases of strong positive correlation, such as that between PDE and OSE.
    \end{itemize}
}

%% file: Sections/6.ttv.tex
\section{Threats To Validity}
\label{sec:ttv}
This section discusses how we prevent and mitigate the threats to the validity of the study, according to \citeauthor{wohlin2012experimentation}~\cite{wohlin2012experimentation}. 

\paragraph{Threats to Construct Validity}
Construct validity concerns arise when there is a mismatch between the conceptual framework and the measurements used~\cite{wohlin2012experimentation}. We used the \textsc{NICHE} dataset~\cite{widyasari2023niche}, which has been previously validated, to mitigate this risk. Additionally, we relied on \textsc{CADOCS}~\cite{voria2022_CADOCS} to compute community smells. Although CADOCS focuses on a subset of community smells, potentially overlooking certain socio-technical challenges in ML-enabled developer communities, it offers valuable insights into the influence of these smells on software development. 

\paragraph{Threats to Conclusion Validity}
Conclusion validity pertains to the reliability of the relationships identified between treatments and outcomes~\cite{wohlin2012experimentation}. A potential threat is the possibility of statistical errors, such as drawing incorrect conclusions due to insufficient sample size or variability in the dataset. To address this, we ensured the dataset was balanced and representative by applying specific selecting criteria established in the literature for similar studies~\cite{catolino2021understanding,tamburri2015open-source}, to create our dataset to conduct the study. This approach enhances the reliability of our findings. Future research could further reduce this risk by replicating the analysis on larger, more diverse datasets and experimenting with alternative statistical methods to validate and extend the study's conclusions.
Another potential threat to the validity of the conclusions concerns the measures used to evaluate our research questions. We adopted a cross-sectional study and calculated prevalence and prevalence odds ratios to quantify the occurrence of community smells and their relationships. Although these measures are well established in similar studies, potential limitations remain. For example, the cross-sectional nature of \textit{RQ\textsubscript{1}} and \textit{RQ\textsubscript{3}} may not capture causal dynamics, and the choice of 3-month time windows in \textit{RQ\textsubscript{2}} may affect trend observations. Future research could explore alternative statistical techniques, such as causal inference methods, to further validate and extend our results.

\paragraph{Threats to External Validity}
External validity concerns the generalizability of the results~\cite{wohlin2012experimentation}. To address this, we built a comprehensive dataset of ML-enabled projects using the validated \textsc{NICHE} dataset~\cite{widyasari2023niche} as a starting point. Our study analyzed 188 projects, employing selection criteria aligned with prior research to enhance relevance and reliability. In addition, projects were examined in 3-month intervals, spanning from inception to 2 years of activity, ensuring longitudinal insights. While this approach supports satisfactory generalizability, future work will expand the sample size and diversity to further validate and generalize the findings.

\paragraph{Threats to Internal Validity}
Threats to internal validity involve factors that could have influenced the study's results~\cite{wohlin2012experimentation}. A potential concern is selection bias from the chosen of \textsc{NICHE} dataset~\cite{widyasari2023niche}, which might limit the generalizability of the findings to all ML-enabled projects. To address this, we employed established selection criteria from the literature to enhance the dataset's representativeness~\cite{catolino2021understanding,tamburri2015open-source}. Another potential issue is measurement bias introduced by the \textsc{CADOCS} tool~\cite{voria2022_CADOCS}, which may not capture all community smells, but only a subset of 10 community smells. 

%% file: Sections/7.conclusion.tex
\section{Conclusion}
\label{sec:conclusion}
This study investigated human factors and behavioral aspects of software engineers and data scientists, analyzing the prevalence, evolution, and correlation of community smells in ML-enabled projects, and emphasizing social interactions between data scientists and software engineers. The study aimed to provide key insights into the challenges of heterogeneous teams developing ML-enabled systems. The study analyzes the prevalence and evolution over time of social behaviors that impact the technical aspects of software, such as dominance behaviors, knowledge silos, and communication breakdowns.
The key contributions are:

\begin{itemize}
    \item \textit{Prevalence analysis}: We established the overall prevalence of community smells, and we see that PDE, SV, and SD are the most dominant in ML-enabled projects.

    \item \textit{Longitudinal analysis}: We analyzed the evolution of community smells over time, revealing the constant persistence of some smells (e.g., PDE) and the variability of others (e.g., UI).

    \item \textit{Correlation analysis}: We identified correlations between community smells, e.g., PDE and OSE, and provided insights into how these smells interact and reinforce each other.
\end{itemize}

The results underscore the socio-technical complexity inherent in ML-enabled projects. 
The high prevalence of community smells PDE underscores the need to address power imbalances within teams. 
Heterogeneous teams composed of data scientists and software engineers should prioritize collaborative workflows and integrated development practices to mitigate these smells.

\textbf{\textit{Future works}} should expand the dataset to analyze a wider range of ML-enabled projects. 
Conducting qualitative studies, such as interviews or case studies, can provide insights into the root causes of community smells and how they manifest in different contexts. 
In-depth longitudinal studies that follow the entire lifecycle of ML projects and analyze the factors that influence the variation of community smells in ML-enabled software.

\section*{Data Availability}

The dataset created, the material and the script used for running the analysis are available in the online appendix~\cite{online_appendix}.